\documentclass{revtex4-2}
\usepackage[utf8]{inputenc}
\usepackage{amsbsy}
\usepackage{amsopn}
\usepackage{amstext}
\usepackage{amsmath,amsthm,amsfonts,amssymb}
\usepackage[mathcal]{eucal}
\usepackage{mathrsfs}
\usepackage{braket}
\usepackage[english]{babel}
\usepackage{color}
\usepackage{esint}
\usepackage{graphicx}
\usepackage{float}
\usepackage{units}
\usepackage{textcomp}
\usepackage{orcidlink}

\DeclareGraphicsExtensions{.png,PNG,.pdf,.PDF}
\usepackage{hyperref}
\usepackage{slashed}
\newcommand{\ie}{\begin{equation}}
\newcommand{\fe}{\end{equation}}
\newcommand{\se}{\begin{eqnarray}}
\newcommand{\ff}{\end{eqnarray}}
\begin{document}

\title{Relativistic and nonrelativistic Landau levels for the noncommutative quantum Hall effect with anomalous magnetic moment in a conical Gödel-type spacetime}
\author{R. R. S. Oliveira\,\orcidlink{0000-0002-6346-0720}}
\email{rubensrso@fisica.ufc.br}
\affiliation{Universidade Federal do Cear\'a (UFC), Departamento de F\'isica,\\ Campus do Pici, Fortaleza - CE, C.P. 6030, 60455-760 - Brazil.}


\date{\today}

\begin{abstract}
In this paper, we analyze the relativistic and nonrelativistic energy spectra (fermionic Landau levels) for the noncommutative quantum Hall effect with anomalous magnetic moment in the conical Gödel-type spacetime in $(2+1)$-dimensions, where such spacetime is the combination of the flat Gödel-type spacetime with a cosmic string (conical gravitational topological defect). To analyze these energy spectra, we start from the noncommutative Dirac equation with minimal and nonminimal couplings in polar coordinates. Using the tetrads formalism, we obtain a second-order differential equation. Next, we solve exactly this differential equation, where we obtain a generalized Laguerre equation, and also a quadratic polynomial equation for the total relativistic energy. By solving this polynomial equation, we obtain the relativistic energy spectrum of the fermion and antifermion. Besides, we also analyze the nonrelativistic limit of the system, where we obtain the nonrelativistic energy spectrum. In both cases (relativistic and nonrelativistic), we discuss in detail the characteristics of each spectrum as well as the influence of all parameters and physical quantities in such spectra. Comparing our problem with other works, we verified that our results generalize several particular cases in the literature.
\end{abstract}
\maketitle

\section{Introduction}

The quantum Hall effect (QHE) \cite{K1,Yoshioka} is the quantized version of the well-known classical Hall effect (CHE) \cite{Hall}, which was discovered by K. von Klitzing $et$ $al$. in 1980 (almost 100 years after the discovery of the CHE \cite{Hall}). For this magnificent discovery, von Klitzing was awarded the 1985 Nobel Prize in Physics. Unlike the CHE, the QHE is ``only'' observed when a two-dimensional electron gas (2DEG) is subjected to low temperatures and strong magnetic fields, consequently, the energy spectrum (Landau levels), the electrical conductivity (Hall conductivity), and the electrical resistivity (Hall resistivity) are then quantized physical quantities. Indeed, for weak fields and ``high'' temperatures, the QHE reduces to the CHE (as it should be). However, in 2007 an exception was observed for the QHE at ``high'' temperatures, that is, it has been shown that in graphene (a single layer of carbon atoms tightly packed in a honeycomb crystal lattice) the QHE can be observed even at room temperature (room-T QHE in graphene) \cite{Novoselov}. In particular, this is due to the highly unusual nature of charged particles in graphene, which behave as massless relativistic particles (massless Dirac fermions), and also move with little scattering under ambient conditions \cite{Novoselov}.

In addition, the QHE has already been studied (theoretically and experimentally) in both nonrelativistic \cite{Yoshioka,Frohlich,Ishikawa1,Thouless,Brand} and relativistic \cite{Haldane,Schakel,Lamata,Miransky,Ishikawa2,Beneventano} quantum mechanics (i.e., in the low and high energy limit), and is also the basis for many modern electronic devices \cite{Popovic,Chien,Lenz}. Currently, the QHE is used as the standard of electrical resistance by metrology laboratories around the world \cite{Jeckelmann}, and recently was studied in problems involving Rabi oscillations \cite{Tran}, Weyl semimetals \cite{Thakurathi}, non-Hermitian and open quantum systems \cite{Yoshida1,Yoshida2}, two-dimensional topological insulator \cite{Shamim}, topological flat bands \cite{Zeng}, etc. It is interesting to mention that in addition to the ordinary, usual or linear QHE, there is also its nonlinear counterpart (latest phenomenon), i.e., the called nonlinear quantum Hall effect (NLQHE) \cite{Sodemann,Nandy,Ma,Kang,Kumar}. Unlike the linear QHE, the NLQHE does not require a broken time-reversal symmetry to exist. Here, we work with the usual QHE.

The concept of noncommutative (NC) spaces (or NC spacetimes) first emerged in 1947 from two papers made by H. S. Snyder on quantized spacetimes \cite{Snyder1,Snyder2}. For Snider, although the Minkowski spacetime is a continuum, this fact is not required by Lorentz invariance. In that way, a model of Lorentz invariant discrete spacetime inspired by quantum mechanics is perfectly possible (in theory), where now the position operators no longer commute with each other (positions are NC variables), i.e.: $[x^{NC}_i,x^{NC}_j]=i\Theta_{ij}$, being $\Theta_{ij}\geq 0$ an antisymmetric (constant) tensor with the dimension of length-squared \cite{Szabo,Douglas,Seiberg}. Later, an even more general concept emerged, the called NC phase space, where now the (linear) momentum operators also no longer commute with each other (moments are also NC variables), i.e.: $[p^{NC}_i,p^{NC}_j]=i\Bar{\Theta}_{ij}$, being $\Bar{\Theta}_{ij}\geq 0$ an antisymmetric (constant) tensor with the dimension of momentum-squared \cite{Szabo,Douglas,Seiberg,Berto,Bastos}. So, from a phenomenological point of view, supposed signatures of NC were investigated in decay of kaons, photon-neutrino interaction, CP violating, etc \cite{Hinchliffe,Melic,Schupp,Abel,Pikovski}. It has been suggested that a possible NC spacetime may arise as one of the effects of quantum gravity \cite{Moffat,Szabo2,Piscicchia1,Piscicchia2}. Also, in order to obtain a more general physical description, the NC (phase) spaces have already been applied in quantum chromodynamics (QCD) \cite{Carlson}, quantum electrodynamics (QED) \cite{Riad}, black holes \cite{Nicolini}, standard model \cite{Melic}, and in various problems of nonrelativistic and relativistic quantum mechanics, including even the QHE \cite{Berto,Bastos,Khordad,Dayi,Dulat,Gamboa,Hassanabadi1,Ho,Oli1}.

In QED, the so-called anomalous magnetic moment (AMM) is a result (or contribution) of the effects of quantized electromagnetic fields (expressed by Feynman diagrams with loops) to the total magnetic dipole moment (MDM) of that fermion (i.e., AMM would be a ``extra internal magnetism-type'') \cite{Greiner1,Greiner2,Geiger}. In particular, the AMM was first discovered by J. S. Schwinger in 1948 (1965 Nobel Prize in Physics together with R. P. Feynman and S. I. Tomonaga) and is symbolically represented by $a=a^{QED}=\frac{\alpha}{2\pi}\approx 0.0011614$ {(one-loop result/correction), where $\alpha\approx\frac{1}{137}$ is the famous fine-structure constant \cite{Schwinger}. Currently, the theoretical value of the AMM (electron) calculated up to the order $\alpha^5$ is given by: $a_e=0.00115965218178(77)$ \cite{Aoyama}, while the experimental value is given by: $a_e=0.00115965218073(28)$ \cite{Hanneke}, i.e., both results agrees by more than 10 significant figures (the most accurately verified prediction in the history of physics). In addition, practically all Dirac fermions (and possibly the neutrinos) have an associated AMM \cite{Greiner1,Greiner2,Geiger,Jegerlehner,Abi,Acciarri,Gutierrez}, where such quantity is also the basis for modeling the Aharonov-Casher effect \cite{Aharonov,Oli2,Oli3}, and the Dirac oscillator \cite{Moshinsky,Martinez,Oli4,Oli5}. Out of curiosity, the main objective of von Klitzing's paper was to show a new method for high-accuracy determination of the fine-structure constant \cite{K1}. Recently, the noncommutative quantum Hall effect (NCQHE) with AMM was studied in three different relativistic scenarios \cite{Oli6}, where their thermodynamic properties were calculated for the flat case \cite{Oli7}.

In General Relativity (GR), one of the most important solutions is certainly the Gödel metric (Gödel spacetime or universe) \cite{Godel}, which represents the first cosmological solution with rotating matter. In particular, this solution is an exact solution of the Einstein field equations, has a negative cosmological constant and a non-null stress-energy tensor, is stationary and spatially homogeneous, possesses cylindrical symmetry, is free of singularities (or horizons), and consists of breaking the causality implying the possibility of closed time-like curves (CTCs), where such CTCs would allow time travels \cite{Deszcz,Gleiser,Barrow}. Also, the Gödel original solution was generalized, among others, by M. J. Rebouças and J. Tiomno \cite{Rebouças}, who found all the spacetime homogeneous solutions of GR with rigid rotation, where such solutions are commonly called Gödel-type solutions, and are characterized by two real parameters: $-\infty<l^2<\infty$ and $\Omega$ (both with the dimensions of length$^{-1}$), where $\Omega\geq 0$ is the vorticity (``rotation'') of the spacetime \cite{Drukker,Carvalho}. With respect to the values of the $l^2$, such a parameter can assume three possible values: a zero value ($l^2=0$), which corresponds to a flat solution (null curvature), a positive value ($l^2>0$), which corresponds to a hyperbolic solution (negative curvature), or a negative value ($l^2<0$), which corresponds to a spherical solution (positive curvature), respectively. However, for $l^2=\Omega^2/2$, we recover the Gödel original solution \cite{Godel}.

Besides, another solution of particular importance in GR is the metric (spacetime) generated by a cosmic string (CS), which is a type of linear gravitational topological defect (still hypothetical) that can arise from a gauge theory with spontaneous symmetry breaking or perhaps may have been formed during phase transitions in the early universe \cite{Carvalho,Oli5,Oli6,Kibble,Vilenkin}. In general, CSs are geometrical deviations from Euclidean geometry characterized by a planar angular deficit: a circle around the outside of a string has a total angle of less than 360° (CSs have a positive curvature). In that way, it says that CSs have a conical singularity along their axis of symmetry ($z$-axis) \cite{Carvalho,Oli5,Oli6,Kibble,Vilenkin}. From an observational point of view, there are some detectors that will try to look for CSs signals, such as the Laser Interferometer Gravitational-Wave Observatory (LIGO) \cite{Cui}, the Laser Interferometer Space Antenna (LISA) \cite{Auclair}, and the North American Nanohertz Observatory for Gravitational Waves (NANOGrav) \cite{Blasi}.

On the other hand, in recent years a brilliant proposal has emerged to work with Gödel-type solutions (metrics) in the presence of a CS, which has resulted in numerous interesting papers on the subject, mainly in the field of relativistic quantum mechanics \cite{Carvalho,Montigny,Garcia,Vitória,Eshghi,Ahmed,Sedaghatnia}. Thus, the general Gödel-type metric (actually the line element) in the presence of a (static) CS in cylindrical coordinates $(t,r,\theta,z)$ with signature $(+,-,-,-)$ is written as follows ($\hslash=c=G=1$)
\begin{equation}\label{metric}
ds^2=\left(dt+\alpha\Omega\frac{\sinh^2lr}{l^2}d\theta\right)^2-dr^2-\alpha^2\frac{\sinh^2 2lr}{4l^2}d\theta^2-dz^2,
\end{equation}
where $0\leq r<\infty$ is the radial coordinate, with $r=\sqrt{x^2+y^2}$, $0\leq \theta\leq 2\pi$ is the angular coordinate, $-\infty<(t,z)<\infty$ are the temporal and axial coordinates, $\Omega$ is called the vorticity parameter, and $\alpha=1-4M$ ($0<\alpha\leq 1$) is the deficit angle or planar angular deficit (also called curvature, topological or disclination parameter) of the conical defect, with $M\geq 0$ being the linear mass density (mass per unit length) of the CS, respectively. So, when $\Omega\to 0$ (null vorticity) we obtain the usual CS spacetime for $\alpha\neq 1$ or the Minkowski spacetime for $\alpha=1$ (usual flat spacetime). Besides, it is well known that in the asymptotic limit $l\to 0$ (with $\alpha\to 1$) the metric above has the same geometry as the Som-Raychaudhuri (SR) metric (modeled only by $\Omega$), i.e., the flat Gödel-type spacetime is also often called the SR spacetime (with limit $l\to 0$ = SR limit), in which it was originally obtained by M. M. Som and A. K. Raychaudhuri in 1968 \cite{Som}. In particular, this metric has appeared several times in string theory \cite{Drukker,Horowitz,Russo1,Russo2}, where also have been reinterpreted as a Gödel-type solution in string theory \cite{Boyda,Harmark}.

This work has as its goal to analyze the relativistic and nonrelativistic energy spectra (fermionic Landau levels) for the NCQHE with AMM in the conical Gödel-type spacetime (flat Gödel-type spacetime with a CS) in $(2+1)$-dimensions ($dz^2=0$). In this case, the conical Gödel-type spacetime is modeled by the parameters $\Omega$ and $\alpha$. So, to analyze such energy spectra, we solve exactly the curved Dirac equation (DE) with minimal and nonminimal couplings in polar coordinates $(t,r,\theta)$, where the two coupling constants are the electric charge and the AMM of the fermion. Besides, we also consider the ``spin'' of the (2D) planar fermion, described by a parameter $s$, called the spin parameter, where $s=+1 (\uparrow)$ is for the spin ``up'', and $s=-1 (\downarrow)$ is for the spin ``down'', respectively. This parameter arose as a result of an exact equivalence between the Aharonov-Bohm effect and the Aharonov-Casher effect (both for spin-1/2 Dirac fermions) \cite{Hagen}. It is interesting to mention that recently the conical Gödel-type spacetime has been studied in the Klein–Gordon \cite{Mustafa1,Mustafa2}.

The structure of this paper is organized as follows. In Sect. \ref{sec2}, we introduce the noncommutative Dirac equation (NCDE) of the system, and using the tetrads formalism (or spin connection), we obtain a second-order differential equation (quadratic NCDE). In Sect. \ref{sec3}, we solve exactly this differential equation through a change of variable and the asymptotic behavior of the equation. Consequently, we obtain a generalized Laguerre equation, and also a quadratic polynomial equation for the total relativistic energy. By solving this polynomial equation, we obtain the relativistic energy spectrum (relativistic Landau levels) for the NCQHE with AMM. In Sect. \ref{sec4}, we analyze the nonrelativistic limit (low-energy limit), where we obtain the nonrelativistic energy spectrum (nonrelativistic Landau levels). In both cases (relativistic and nonrelativistic), we discuss in detail the characteristics of each spectrum as well as the influence of all parameters and physical quantities in such spectra. We also compared our problem with other works, where we verified that our results generalize several particular cases in the literature. Finally, in Sect. \ref{sec5} we present our conclusions. Here, we use the natural units $(\hslash=c=G=1)$ and the spacetime with signature $(+,-,-)$.

\section{The noncommutative Dirac equation in the conical Gödel-type spacetime \label{sec2}}

The tensorial DE with minimal and nonminimal couplings in a generic curved spacetime is given by the following expression (in polar coordinates) \cite{Oli6,Greiner2,Lawrie,B1,B2,Maluf}
\begin{equation}\label{dirac1}
\left\{\gamma^\mu(x)[P_\mu(x)-qA_\mu (x)]+\frac{a\vert q \vert}{4m_0}\sigma^{\mu\nu}(x)F_{\mu\nu}(x)-m_0\right\}\psi(t,r,\theta)=0, \ \ (\mu,\nu=t,r,\theta),
\end{equation}
where $\gamma^{\mu}(x)=e^\mu_{\ a}(x)\gamma^a$ are the curved gamma matrices, with $\gamma^a=(\gamma^0,\gamma^1,\gamma^2)$ being the usual or flat gamma matrices (in Cartesian coordinates) and $e^\mu_{\ a}(x)$ are the inverse tetrads (sometimes simply called tetrads), $P_\mu(x)=i\nabla_\mu (x)=i[\partial_\mu+\Gamma_\mu (x)]$ is the curved momentum operator, $\nabla_\mu(x)$ is the covariant derivative (operator), with $\partial_\mu=(\partial_t,\partial_r,\partial_\theta)$ being the usual partial derivatives, $\Gamma_\mu(x)=-\frac{i}{4}\omega_{ab\mu}(x)\sigma^{ab}$ is the spinorial connection (or spinor affine connection), with $\omega_{ab\mu}(x)$ being the spin connection and $\sigma^{ab}=\frac{i}{2}[\gamma^a,\gamma^b]=i\gamma^a\gamma^b$ ($a\neq b$) is a flat ``Dirac'' antisymmetric tensor, whose curved counterpart (curved ``Dirac'' antisymmetric tensor) is given by $\sigma^{\mu\nu}(x)=\frac{i}{2}[\gamma^\mu(x),\gamma^\nu(x)]=i\gamma^\mu(x)\gamma^\nu(x)$ ($\mu\neq \nu$), $F_{\mu\nu}(x)=\partial_\mu A_\nu (x)-\partial_\nu A_\mu (x)$ (or yet $F_{\mu\nu}(x)=e^a_{\ \ \mu}(x)e^b_{\ \ \nu}(x)F_{ab}$) is the curved electromagnetic field tensor (curved field strength tensor), with $F_{ab}$ being the flat electromagnetic field tensor and $A_\mu (x)=e^b_{\ \mu}(x)A_{b}$ is the curved electromagnetic potential (curved gauge field), being $e^b_{\ \mu}(x)$ the tetrads, $\psi=\ket{\psi}=e^{\frac{i\theta\Sigma^3}{2}}\Psi_D$ is the two-component curvilinear spinor, where $\Psi_D\in\mathbb{C}^2$ is the original Dirac spinor, and the (invariant) physical quantities symbolized by $q=\pm e$, $a>0$, and $m_0>0$, are the electric charge, the AMM, and the rest mass of the Dirac fermion, respectively. Here, the Latin indices $(a, b, c, \ldots)$ are used to label the local coordinates system (local reference frame or the Minkowski spacetime) while the Greek indices $(\mu, \nu, \alpha, \ldots)$ is for the general coordinates system (general reference frame or the curved spacetime).

Explicitly, Eq. \eqref{dirac1} can be rewritten as
\cite{Oli6}
\begin{equation}\label{dirac2}
\left\{\gamma^t(x)P_t(x)+\gamma^{i}(x)\left[P_{i}(x)+\frac{qB}{2}\epsilon_{ij}x^{j}(x)\right]-2\mu_m\vec{S}\cdot\vec{B}-m_0\right\}\psi=0, \ \ (i,j=r,\theta),
\end{equation}
where we use $A_\mu(x)=(0,A_{i}(x))$, with $A_{i}(x)=-\frac{B}{2}\epsilon_{ij}x^{j}(x)$ being the spatial components of the curved vector potential and $B=B_3=B_z>0$ is the strength (module) of an external uniform magnetic field along the $z$-axis, given by $\Vec{B}=B\Vec{e}_z$ (is the magnetic field of the QHE), and we also use the fact that $\sigma^{\mu\nu}(x)F_{\mu\nu}(x)=\sigma^{ab}F_{ab}=-4\Vec{S}\cdot\Vec{B}$ (``spin-magnetic field coupling''), where $\Vec{S}=\frac{1}{2}\Vec{\Sigma}$ is the spin vector (operator), and $\mu_m\equiv a\mu_B$ is the anomalous MDM, or simply the MDM (of the relativistic fermion), with $\mu_B=\frac{\vert q \vert}{2m_0}$ being the famous Bohr magneton. It is important to mention that this MDM (originated by the AMM) not must be confused with the two MDMs of nonrelativistic quantum mechanics (nonrelativistic limit of the DE$_{(3+1)D}$), that is the orbital MDM (originated by the orbital angular momentum) and the spin MDM (originated by the spin angular momentum) \cite{Greiner2,Griffiths}.

So, in a NC phase space, we have the following NCDE
\begin{equation}\label{dirac3}
\left\{\gamma^t(x)P_t(x)+\gamma^{i}(x)\left[P^{NC}_{i}(x)+\frac{qB}{2}\epsilon_{ij}(x^{j}(x))^{NC}\right]-2\mu_m\vec{S}\cdot\vec{B}-m_0\right\}\star\psi=0,
\end{equation}
where the symbol $\star$ is the called star product or Moyal product, and now the product of two arbitrary functions $F$ and $G$ is given by
\begin{equation}\label{starproduct}
F\star G\equiv Fe^{(i/2)(\overleftarrow{\partial}_{x_i}\Theta_{ij}\overrightarrow{\partial}_{x_j})}G=Fe^{(i\Theta/2)(\overleftarrow{\partial}_{x}\overrightarrow{\partial}_{y}-\overleftarrow{\partial}_{y}\overrightarrow{\partial}_{x})}G.
\end{equation}

However, knowing that the NC operators $P^{NC}_{i}(x)$ and $(x^{j}(x))^{NC}$ can be written as follows \cite{Oli6}
\begin{equation}\label{operators}
(x^{j}(x))^{NC}=x^{j}(x)-\frac{1}{2}\Theta\epsilon^{jl}P_{l}(x), \ \ P^{NC}_{i} (x)=P_{i}(x)+\frac{1}{2}\Bar{\Theta}\epsilon_{ij}x^{j}(x),
\end{equation}
Eq. (\ref{dirac3}) becomes
\begin{equation}\label{dirac4}
\left\{\gamma^t(x)P_t(x)+\gamma^{i}(x)\left[\tau P_{i}(x)-q\lambda A_{i}(x)\right]-2\mu_m\vec{S}\cdot\vec{B}-m_0\right\}\psi^{NC}=0,
\end{equation}
or yet
\begin{eqnarray}\label{dirac5}
\left[i\gamma^t(x)\partial_t+i\tau\gamma^r(x)\partial_r+i\tau\gamma^\theta(x)\partial_\theta+e\lambda\gamma^{\theta}(x)A_{\theta}(x)-\mu_m\Sigma^3 B-m_0
\right]\psi^{NC} 
 \nonumber \\
    +i[\gamma^t(x)\Gamma_t(x)+\tau\gamma^{r}(x)\Gamma_{r}(x)+\tau\gamma^{\theta}(x)\Gamma_{\theta}(x)]\psi^{NC}=0,
\end{eqnarray}
where $\tau\equiv (1-eB\Theta/4)>0$ and $\lambda\equiv(1-\Bar{\Theta}/eB)>0$ are two real parameters, $\psi^{NC}$ is the NC spinor, and we also consider $q=-e$ ($e>0$). For simplicity, here we consider these two parameters to be positives \cite{Oli6} and, consequently, this restricts (limits) the possible values (or allowed values) of $\Theta$ and $\Bar{\Theta}$ (for a given arbitrary value or any of $eB$), i.e., $\Theta$ and $\Bar{\Theta}$ must satisfy the conditions: $0\leq \Theta<4/eB$ and $0\leq\Bar{\Theta}<eB$. Furthermore, as we will soon see, the product $\tau\lambda$ will appear in the energy spectra, where such quantity will result in a quadratic polynomial inequality for $B$ and, consequently, will restrict (limit) the possible values (or allowed values) of $B$ (for a given arbitrary value or any of $e$, $\Theta$ and $\Bar{\Theta}$).

Now, we will focus our attention on the line element of the conical Gödel-type spacetime in $(2+1)$-dimensions as well as on the form of the metric, tetrads, curved gamma matrices, and spinorial and spin connections. So, at the limit $l\to 0$ ($l^2=0$), the line element \eqref{metric} is reduced to the following line element of the conical Gödel-type spacetime \cite{Carvalho}
\begin{equation}\label{lineelement1}
ds^2=g_{\mu\nu}(x)dx^\mu dx^\nu=(dt+\alpha\Omega r^2 d\theta)^2-dr^2-\alpha^2 r^2d\theta^2,
\end{equation}
where $g_{\mu\nu}(x)$ is the curved metric tensor (or simply the curved metric) at the point $x$, whose inverse is given $g^{\mu\nu}(x)$, where both are written as
\begin{equation}\label{metric1}
g_{\mu\nu}(x)=\left(\begin{array}{ccc}
1 & \ 0 & \alpha\Omega r^2 \\
0 & -1 &  0 \\
\alpha\Omega r^2 & \ 0 & \alpha^2\Omega^2 r^4-\alpha^2 r^2
\end{array}\right), \ \
g^{\mu\nu}(x)=\left(\begin{array}{ccc}
1-\Omega^2 r^2 & 0 & \frac{\Omega}{\alpha}\\
0 & -1 & 0\\
\frac{\Omega}{\alpha} & 0 & -\frac{1}{\alpha^2 r^{2}}
\end{array}\right),
\end{equation}

Thus, with the line element well defined (or metric well defined), given by the expression \eqref{lineelement1}, we now need to build a local reference frame where the observer will be placed (i.e., laboratory frame). Consequently, it is in this local frame that we can then define the gamma matrices (or the spinor) in a curved spacetime \cite{Oli6,B1,B2,Lawrie,Maluf}. In particular, using the tetrad formalism (of GR), it is perfectly possible to achieve this objective \cite{Oli6,B1,B2,Lawrie,Maluf} (for a more complete analysis of the tetrad formalism we recommend Refs. \cite{Lawrie,Maluf}). According to this formalism, a curved spacetime can be introduced point to point with a flat spacetime through objects of the type $e^a_{\ \mu}(x)$, which are called tetrads (square matrices), and which together with their inverses, given by $e^\mu_{\ a}(x)$, satisfy the following relations \cite{Oli6,B1,B2,Lawrie,Maluf}: $\hat{\theta}^a=e^a_{\ \mu}(x)dx^\mu$ and $dx^\mu=e_{\ a}^\mu(x)\hat{\theta}^a$, where $\hat{\theta}^a$ ($a=0,1,2$) is a quantity called noncoordinate basis (not to be confused with the angular coordinate $\theta$). Furthermore, the tetrads and their inverses must also satisfy the following relations \cite{Oli6,B1,B2,Lawrie,Maluf}
\begin{eqnarray}\label{metric2}
&& g_{\mu\nu}(x)=e^a_{\ \mu}(x)e^b_{\ \nu}(x)\eta_{ab},
\nonumber\\
&& g^{\mu\nu}(x)=e^\mu_{\ a}(x)e^\nu_{\ b}(x)\eta^{ab},
\nonumber\\
&& g^{\mu\sigma}(x)g_{\nu\sigma}(x)=\delta^\mu_{\ \nu},
\end{eqnarray}
where $\eta_{ab}=\eta^{ab}=(\eta_{ab})^{-1}=$diag$(1,-1,-1)$ is the Cartesian Minkowski metric tensor (or simply the flat metric or Lorentzian metric), in which must satisfy
\begin{eqnarray}\label{metric3}
&& \eta_{ab}=e^\mu_{\ a}(x)e^\nu_{\ b}(x)g_{\mu\nu}(x),
\nonumber\\
&& \eta^{ab}=e^a_{\ \mu}(x)e^b_{\ \nu}(x)g^{\mu\nu}(x),
\nonumber\\
&& \eta^{ac}\eta_{cb}=\delta^a_{\ b},
\end{eqnarray}
with $\delta^a_{\ b}=\delta^\mu_{\ \nu}=I_{3\times 3}$=diag$(+1,+1,+1)$ being the $(2+1)$-dimensional Kronecker delta, which is exactly the $3\times 3$ identity matrix (unit matrix), and $e^a_{\ \mu}(x)$ and $e^b_{\ \nu}(x)$ are the inverse tetrads of $e^\mu_{\ a}(x)$ and $e^\nu_{\ b}(x)$, in which satisfy
\begin{equation}\label{delta}
 e^a_{\ \mu}(x)e^\mu_{\ b}(x)=\delta^a_{\ b}, \ \ e^a_{\ \nu}(x)e^\mu_{\ a}(x)=\delta^\mu_{\ \nu}.
\end{equation}

So, through the tetrad formalism, we can rewrite the line element \eqref{lineelement1} in terms of the noncoordinate basis, such as
\begin{equation}\label{lineelement2}
ds^2=\eta_{ab}\hat{\theta}^a\otimes\hat{\theta}^b=(\hat{\theta}^0)^2-(\hat{\theta}^1)^2-(\hat{\theta}^2)^2,
\end{equation}
where the components of $\hat{\theta}^a$ are given by (i.e., well-defined noncoordinated basis)
\begin{equation}\label{bases}
\hat{\theta}^0=dt+\alpha\Omega r^2 d\theta, \ \ \hat{\theta}^1=dr, \ \ \hat{\theta}^2=\alpha r d\theta, \ \ (dx^t=dt, \ dx^r=dr, \  dx^\theta=d\theta).
\end{equation}

Therefore, using the relations $\hat{\theta}^a=e^a_{\ \mu}(x)dx^\mu$ and $dx^\mu=e_{\ a}^\mu(x)\hat{\theta}^a$, implies that the tetrads and their inverses take the following form
\begin{equation}\label{tetrads}
e^{\mu}_{\ a}(x)=\left(
\begin{array}{ccc}
 1 & 0 & -\Omega r \\
 0 & 1 & 0 \\
 0 & 0 & \frac{1}{\alpha r} \\
\end{array}
\right), \ \
e^{a}_{\ \mu}(x)=\left(\begin{array}{ccc}
1 & 0 & \alpha\Omega r^2 \\
0 & 1 & 0 \\
0 & 0 & \alpha r
\end{array}\right).
\end{equation}

Consequently, the curved gamma matrices are written as
\begin{eqnarray}\label{gammamatrices}
&& \gamma^t(x)=\gamma^0-\Omega r\gamma^2,
\nonumber\\
&& \gamma^r(x)=\gamma^1,
\nonumber\\
&& \gamma^\theta(x)=\frac{1}{\alpha r}\gamma^2.
\end{eqnarray}

On the other hand, to find the form of the spinorial connection we first need to find the form of the spin connection (antisymmetric tensor in its internal indices $a$ and $b$). According to Ref. \cite{Lawrie}, this spin connection is defined as follows (torsion-free) 
\begin{equation}\label{spinconnection1}
\omega_{ab\mu}(x)=-\omega_{ba\mu}(x)=\eta_{ac}e^c_{\ \nu}(x)\left[e^\sigma_{\ b}(x)\Gamma^\nu_{\ \mu\sigma}(x)+\partial_\mu e^\nu_{\ b}(x)\right], 
\end{equation}
where $\Gamma^\nu_{\ \mu\sigma}(x)$ are the Christoffel symbols of the second type (symmetric tensor), and written as
\begin{equation}\label{Christoffel}
\Gamma^\nu_{\ \mu\sigma}(x)=\frac{1}{2}g^{\nu\lambda}(x)\left[\partial_\mu g_{\lambda\sigma}(x)+\partial_\sigma g_{\lambda\mu}(x)-\partial_\lambda g_{\mu\sigma}(x)\right]. 
\end{equation}

Explicitly, the non-null components of the Christoffel symbols are given by
\begin{eqnarray}\label{symbols}
&& \Gamma^{t}_{\ tr}=\Gamma^{t}_{\ rt}=\Omega^2 r,
\nonumber\\
&& \Gamma^{t}_{\ r\theta}=\Gamma^{t}_{\ \theta r}=\alpha\Omega^3 r^3,
\nonumber\\
&& \Gamma^{r}_{\ t \theta}=\Gamma^{r}_{\ \theta t}=\alpha\Omega r,
\nonumber\\
&& \Gamma^{r}_{\ \theta\theta}=2\alpha^2\Omega^2 r^3-\alpha^2 r,
\nonumber\\
&& \Gamma^{\theta}_{\ tr}=\Gamma^{\theta}_{\ rt}=-\frac{\Omega}{\alpha r},
\nonumber\\
&& \Gamma^{\theta}_{\ r\theta}=\Gamma^{\theta}_{\ \theta r}=\frac{1-\Omega^2 r^2}{r}.
\end{eqnarray}

Consequently, the spin connection is written as (non-null components)
\begin{eqnarray}\label{spinconnection2}
&& \omega_{12t}=-\omega_{21t}=\Omega,
\nonumber\\
&& \omega_{02r}=-\omega_{20r}=\Omega,
\nonumber\\
&& \omega_{01\theta}=-\omega_{10\theta}=-\alpha\Omega r,
\nonumber\\
&& \omega_{12\theta}=-\omega_{21\theta}=\alpha\Omega^2 r^2-\alpha,
\end{eqnarray}
where implies in the following spinorial connection (non-null components)
\begin{eqnarray}\label{spinorialconnection}
&& \Gamma_t=-\frac{\Omega}{2}\gamma^1\gamma^2,
\nonumber\\
&& \Gamma_r=-\frac{\Omega}{2}\gamma^0\gamma^2,
\nonumber\\
&& \Gamma_\theta=\frac{(-\alpha\Omega^2 r^2+\alpha)\gamma^1\gamma^2+(\alpha\Omega r)\gamma^0\gamma^1}{2}.
\end{eqnarray}

Then, using the expressions \eqref{gammamatrices} and \eqref{spinorialconnection}, we obtain the following contribution of the spinorial connection (or spin)
\begin{equation}\label{contributionofthespinorialconnection}
[\gamma^t(x)\Gamma_t(x)+\tau\gamma^{r}(x)\Gamma_{r}(x)+\tau\gamma^{\theta}(x)\Gamma_{\theta}(x)]=\frac{(2\tau-1)\Omega}{2}\gamma^{0}\gamma^1\gamma^2+\frac{(1-\tau)\Omega^2 r^2+\tau}{2 r}\gamma^{1}.
\end{equation}

Therefore, using now the expression above in \eqref{dirac5}, we have the following NCDE in a conical Gödel-type spacetime
\begin{equation}\label{dirac6}
\left[(\gamma^0-\Omega r\gamma^2)i\partial_t+i\tau\gamma^1\partial_r+i\frac{\tau}{\alpha r}\gamma^2\partial_\theta-\frac{\lambda m_0\omega_c}{2}r\gamma^2+\frac{(2\tau-1)\Omega}{2}i\gamma^{0}\gamma^1\gamma^2+\frac{(1-\tau)\Omega^2 r^2+\tau}{2 r}i\gamma^{1}-(\mu_m\Sigma^3 B+m_0)\right]\psi^{NC}=0,
\end{equation}
where we use the fact that $\gamma^{\theta}(x)A_{\theta}(x)=-\frac{1}{2}Br\gamma^2$, with $A_{\theta}(x)$ given by $A_{\theta}(x)=\alpha r A_2=-\alpha r A_\theta=-\frac{1}{2}B\alpha r^2$ (a potential vector quadratic in the radial coordinate) \cite{Cunha1,Cunha2,Oli5}, being $A_\theta=\frac{1}{2}Br$ (Landau symmetric gauge) the angular component of $A_b=\eta_{bc}A^c=(A_0,A_1,A_2)=(0,0,-A_\theta)$, and $\omega_c=\frac{eB}{m_0}\geq 0$ is the famous cyclotron frequency (an angular velocity). In particular, the product $\gamma^\theta (x)A_\theta (x)$ could also be written as: $\gamma^\theta (x)A_\theta (x)=e^\theta_{\ 2}(x)e^2_{\ \theta}(x)\gamma^2 A_2=\delta^2_{\ 2}\gamma^2 A_2=\gamma^2 A_2=-\gamma^2A_{\theta}$, with $A_{\theta}=\frac{1}{2}Br$ (a potential vector linear in the radial coordinate), that is, we have the product in the Minkowski spacetime (in polar coordinates) already done a similarity transformation (this explains the existence of exponential $e^{\frac{i\theta\Sigma^3}{2}}$ in the spinor $\psi$, where such exponential has the purpose of converter $\gamma^\theta$ in $\gamma^2$ and $\gamma^r$ in $\gamma^1$) \cite{Oli6}.

In addition, here we assume that our system is a stationary quantum system (``stationary Hall states''); consequently, the two-component NC spinor can be written as follows \cite{Oli6}
\begin{equation}\label{spinor}
\psi^{NC}(t,r,\theta)=\frac{e^{i(m_j\theta-Et)}}{\sqrt{2\pi}}\phi^{NC}, \ \ \phi^{NC}=\left(
           \begin{array}{c}
            f_+(r) \\
            if_-(r) \\
           \end{array}
         \right),
\end{equation}
where $f_+(r)$ and $f_-(r)$ are real radial functions ($f_+\neq f_-$), $E$ is the relativistic total energy, and $m_j=\pm\frac{1}{2},\pm\frac{3}{2},\pm\frac {5}{2},\ldots$ is the total (angular momentum) magnetic quantum number.

So, applying the spinor \eqref{spinor} in \eqref{dirac6}, we have the following time-independent NCDE (or stationary NCDE)
\begin{equation}\label{dirac8}
\left\{i\tau\gamma^1\left(\partial_r+\frac{1}{2r}+\frac{(1-\tau)\Omega^2}{2\tau}r\right)-\gamma^2\left[\frac{\tau m_j}{\alpha r}+\left(\frac{\lambda m_0\omega_c}{2}+\Omega E\right)r\right]-[\mu_m\Sigma^3 B+m_0-\gamma^0 E-(2\tau-1)\gamma^0\Vec{S}\cdot\Vec{\Omega}]\right\}\phi^{NC}=0,
\end{equation}
where the term $\Vec{S}\cdot\Vec{\Omega}=\frac{1}{2}\Omega\Sigma^3$ is a ``spin-vorticity coupling'', with $\Vec{S}=\frac{1}{2}\Vec{\Sigma}$ (spin vector) and $\Sigma^3=\Sigma_3=i\gamma^1\gamma^2$. In particular, this coupling is analogous to the spin-rotation coupling of Dirac fermions in rotating reference frames \cite{Oli2,Oli5,Oli6}.

However, in order to obtain a generalized Laguerre equation (we will see this later), we can also simplify Eq.\eqref{dirac8} by eliminating the third term of this equation, where such simplification can be done by redefining the spinor $\phi^{NC} $ as follows (as we will see shortly, this will still satisfy the boundary conditions of the problem)
\begin{equation}\label{redefining}
 \phi^{NC}\to\bar{\phi}^{NC}\equiv e^{-\frac{(1-\tau)\Omega^2 r^2}{4\tau}}\phi^{NC}.
\end{equation}

Therefore, Eq. \eqref{dirac8} becomes
\begin{equation}\label{dirac9}
\left\{i\tau\gamma^1\left(\partial_r+\frac{1}{2r}\right)-\gamma^2\left[\frac{\tau m_j}{\alpha r}+\left(\frac{\lambda m_0\omega_c}{2}+\Omega E\right)r\right]-[\mu_m\Sigma^3 B+m_0-\gamma^0 E-(2\tau-1)\gamma^0\Vec{S}\cdot\Vec{\Omega}]\right\}\bar{\phi}^{NC}=0.
\end{equation}

On the other hand, since we are working in a $(2+1)$-dimensional spacetime, the flat gamma matrices are written in the form $\gamma^a=\eta^{ab}\gamma_b=(\gamma^0,\gamma^1,\gamma^2)=(\gamma_0,-\gamma_1,-\gamma_2)$, with $\gamma_1=\sigma_3\sigma_1=i\sigma_2$,  $\gamma_2=s\sigma_3\sigma_2=-is\sigma_1$ ($s=\pm 1$ is the spin parameter), and $\gamma^0=\Sigma^3=\sigma_3$ \cite{Oli6}. Besides, knowing that the Pauli matrices are written as
\begin{equation}\label{Paulimatrices}
\sigma_1=\left(
    \begin{array}{cc}
      0\ &  1 \\
      1\ & 0 \\
    \end{array}
  \right), \ \  \sigma_2=\left(
    \begin{array}{cc}
      0 & -i  \\
      i & \ 0 \\
    \end{array}
  \right), \ \  \sigma_3=\left(
    \begin{array}{cc}
      1 & \ 0 \\
      0 & -1 \\
    \end{array}
  \right),
\end{equation}
implies that we can obtain from \eqref{dirac9} two coupled first-order differential equations for $f_+(r)$ and $f_-(r)$, given by
\begin{equation}\label{dirac10}
A_{-}f_+(r)=\left[\frac{d}{dr}+sm_0\omega r+\frac{s}{\alpha r}\left(m_j+\frac{s\alpha}{2}\right)\right]f_-(r),
\end{equation}
\begin{equation}\label{dirac11}
A_{+}f_-(r)=\left[\frac{d}{dr}-sm_0\omega r-\frac{s}{\alpha r}\left(m_j-\frac{s\alpha}{2}\right)\right]f_+(r),
\end{equation}
where
\begin{equation}\label{frequency}
A_\pm\equiv\frac{\left[\left(m_0-\frac{(2\tau-1)\Omega}{2}\right)\pm(E-\mu_m B)\right]}{\tau}, \ \ \omega=\omega_{eff}=\omega_{total}\equiv\left(\frac{\lambda\omega_c}{2\tau}+\frac{\Omega E}{m_0\tau}\right)\geq 0,\ (\Omega E=\vert \Omega E\vert\geq 0),
\end{equation}
being $\omega$ an effective or total angular frequency.

Therefore, substituting \eqref{dirac11} into \eqref{dirac10}, and then \eqref{dirac10} into \eqref{dirac11}, i.e., decoupling the equations, we obtain the following second-order differential equation (“quadratic NCDE” or “second-order NCDE”) for the NCQHE with MMA in a conical Gödel-type spacetime
\begin{equation}\label{dirac12}
\left[\frac{d^2}{dr^2}+\frac{1}{r}\frac{d}{dr}-\frac{\gamma^2_u}{\alpha^2 r^2}-(m_0\omega r)^2+E_u\right]f_u(r)=0,
\end{equation}
where we define
\begin{eqnarray}\label{dirac13}
&& \gamma_u\equiv\left(m_j-\frac{su\alpha}{2}\right)\gtrless 0,
\nonumber\\
&& E_u\equiv\frac{(E-\mu_m B)^2-M_0^2}{\tau^2}-\frac{2m_0\omega}{\alpha}\left(m_j+\frac{su\alpha}{2}\right),
\nonumber\\
&& M_0=M_{eff}=M^{NC}_\Omega\equiv\left(m_0-\frac{(2\tau-1)}{2}\Omega\right)>0,
\end{eqnarray}
with $M_0$ being a ``NC Gödel or effective mass'' and the parameter $u=\pm 1$ (spinorial parameter) was introduced to represent the two components of the spinor ($u=+1$ is for the upper component and $u=-1$ is for the lower component), respectively.

\section{The relativistic energy spectrum (relativistic Landau levels)\label{sec3}}

To solve exactly Eq. \eqref{dirac12}, let us first introduce a new (dimensionless) variable, given by: $\rho=m_0\omega r^2\geq 0$. Thus, through a change of variable, we have
\begin{equation}\label{dirac14}
\left[\rho\frac{d^{2}}{d\rho^{2}}+\frac{d}{d\rho}-\frac{\gamma^{2}_u}{4\alpha^2\rho}-\frac{\rho}{4}+\frac{E_u}{4m_0\omega}\right]f_u(\rho)=0.
\end{equation}

Now, analyzing the asymptotic (limit) behavior of Eq. \eqref{dirac14} for $\rho\to 0$ and $\rho\to\infty$, we get a regular solution to this equation given by
\begin{equation}\label{dirac15}
f_u(\rho)=C_u\rho^{\frac{\vert\gamma_u\vert}{2\alpha}}e^{-\frac{\rho}{2}}F_u(\rho), \ \ (\vert\gamma_u\vert>0),
\end{equation}
where $C_u>0$ are normalization constants, $F_u(\rho)$ are unknown functions to be determined, and $f_u(\rho)$ must satisfy the following boundary conditions to be a normalizable solution (finite or physically acceptable solution)
\begin{equation}\label{conditions} 
f_u(\rho\to 0)=f_u(\rho\to\infty)=0.
\end{equation}

Therefore, we clearly see that the redefinition given in \eqref{redefining} also satisfies the boundary conditions. So, the two real exponentials for the Dirac spinor can be combined into one, given by: $e^{-\frac{\bar{\omega}r^2}{2}}$, with $\bar{\omega}\equiv\left((1-\tau)\frac{\Omega^2}{2\tau}+m_0\omega\right)\geq 0$ (otherwise, i.e., $\bar{\omega}<0$, it does not satisfy the boundary conditions).

So, substituting \eqref{dirac15} in \eqref{dirac14}, we have a second-order differential equation for $F_u(\rho)$ given by
\begin{equation}\label{dirac16}
\left[\rho\frac{d^{2}}{d\rho^{2}}+(\vert\bar{\gamma}_u\vert-\rho)\frac{d}{d\rho}-\Bar{E}_u\right]F_u(\rho)=0,
\end{equation}
where we define
\begin{equation}\label{define}
\vert\bar{\gamma}_u\vert\equiv\frac{\vert \gamma_u\vert}{\alpha}+1, \ \ \Bar{E}_u\equiv\frac{\vert\bar{\gamma}_u\vert}{2}-\frac{E_u}{4m_0\omega}.
\end{equation}

It is not difficult to note that Eq. \eqref{dirac16} is the well-known generalized Laguerre equation, whose solutions are the generalized Laguerre polynomials, written as $F_u(\rho)=L^{\frac{\vert \gamma_u\vert}{\alpha}}_n(\rho)$  \cite{Greiner2}. Consequently, the quantity $\Bar{E}_u$ must be equal to a non-positive integer, i.e., $\Bar{E}_u=-n$ (a quantization condition), where $n=n_r=0,1,2,\ldots$ is the (radial) quantum number (also called the Landau level index). Therefore, from this condition, we obtain as a result the following quadratic polynomial equation for the total relativistic energy $E$
\begin{equation}\label{dirac17}
E^2-2(\mu_m B+2\tau N\Omega)E+\left[(\mu_m B)^2-M^2_0-2m_0\tau\lambda\omega_c N\right]=0,
\end{equation}
where
\begin{equation}\label{N1}
N=N_{eff}=N_{total}\equiv\left(n+\frac{1}{2}+\frac{\Big|m_j-\frac{su\alpha}{2}\Big|+\left(m_j+\frac{su\alpha}{2}\right)}{2\alpha}\right)\geq 0,
\end{equation}
with $N$ being a total or effective quantum number. In particular, what will tell us whether $N$ is greater or equal to zero are the values of $m_j$, $s$, and $u$ (we will see this shortly).

Therefore, solving the polynomial equation \eqref{dirac17}, we obtain as a result the following relativistic energy spectrum (relativistic Landau levels) for the NCQHE with MMA in a conical Gödel-type spacetime
\begin{equation}\label{spectra}
E^\sigma_{n,m_j,s,u}=E_m+ E_\Omega^{NC}+\sigma\sqrt{(E_m+E_\Omega^{NC})^2-E^2_m+M_{0}^2+\epsilon^{NC}_N},
\end{equation}
where we define
\begin{equation}\label{define}
E_{m}\equiv\mu_m B=\frac{a\omega_c}{2} \ \left(\mu_m=\frac{ae}{2m_0}\right), \ \ E_\Omega^{NC}\equiv 2\tau\Omega N, \ \ \epsilon^{NC}_N\equiv 2m_0\omega_c^{NC}N, \ \ \omega_c^{NC}\equiv\tau\lambda\omega_c,
\end{equation}
with $\sigma=\pm 1$ being a parameter (``energy parameter'') which represents the positive energy states/solutions or the particle states (with $E_{electron}=E^+>0$) and the negative energy states/solutions or the antiparticle states (with $E_{positron}=-E^-=\vert E^-\vert>0$ \cite{Greiner2}), in which both particle and antiparticle have practically the same anomalous magnetic potential energy (or simply magnetic energy), given by $E_m$ ($E_m^+\simeq E_m^-$) \cite{Van}. As we can see, this energy arises from the interaction of the MMA (or MDM) with the external magnetic field $B$ (is not quantized because it does not depend on $N$, i.e., is a continuous energy as a function of $B$, or even $\omega_c$). In addition, the quantity $E_\Omega^{NC}$ is a ``NC Gödel spectrum'' (is quantized due to $N$), and arises due to the existence of the vorticity parameter $\Omega$ (for $\Omega=0$ this energy is completely null), and the quantity $\epsilon_N^{NC}$ is a ``NC cyclotron spectrum'' (also quantized due to $N$) since it depends directly on the NC cyclotron frequency $\omega_c^{NC}$ (i.e., is the original cyclotron frequency modified by NC phase space). As we can see, this energy arises from the interaction of the electrical charge of the particle/antiparticle with the external magnetic field $B$ (the origin of Landau levels comes from this, i.e., the called Landau quantization refers to the quantization of the orbits of charged particles in uniform magnetic fields). In particular, the spectra \eqref{spectra} also do not have a well-defined degeneracy, in other words, the function of the parameter $\alpha$ is to break the degeneracy of the spectra (i.e., the presence of a CS modifies the energy levels and breaks the degeneracy of the eigenvalues) \cite{Carvalho}.

So, although the magnetic energy $E_m$ is practically equal for both particle and antiparticle, the total energy is not, i.e., we have $\vert E^+\vert >\vert E^-\vert$ (this arises because $E_m$ and $E_\Omega^{NC}$ are added to the square root energy for the case of the particle and subtracted for the case of the antiparticle). In this case, the spectra are said to be asymmetric, where the total energy of the particle is always greater than that of the antiparticle. In particular, this asymmetry arises due to energies $E_m$ and $E^{NC}_\Omega$, i.e., these energies break the symmetry of the spectrum since for $E_m=E^{NC}_\Omega=0$, or $\mu_m=\Omega=0$, we have a symmetric spectrum (such as $\vert E_+\vert=\vert E_-\vert$). Furthermore, the spectrum \eqref{spectra} carries a mass mixed with vorticity (in $M_0$), which is a direct consequence of spin-vorticity coupling. In particular, this is completely different from what occurs in the bosonic case of spin-0 particles (spinless particles), where $M_0=M$ is simply the (rest) mass of the particle (in fact, as Klein-Gordon bosons have no spin, it means that there is no spin-vorticity coupling, as happens for the Dirac fermions) \cite{Carvalho}.

On the other hand, the presence of NC parameters $\Theta$ and $\Bar{\Theta}$ (in $\tau$ and $\lambda$) allow defining two ``NC angular frequencies'', or “cyclotron-type frequencies'' (has dimension of $\omega_c$), defined as: $\omega_\Theta\equiv\frac{4}{m_0\Theta}$ and $\omega_{\bar{\Theta}}\equiv\frac{\bar{\Theta}}{m_0}$, where $\omega_\Theta$ is the NC frequency of position (decreases with increasing $\Theta$) and $\omega_{\bar{\Theta}}$ is the NC frequency of momentum (increases with increasing $\bar{\Theta}$), respectively. With this, we can rewrite the parameters $\tau$ and $\lambda$ as: $\tau=(1-\omega_c/\omega_\Theta)$ and $\lambda=(1-\omega_{\bar{\Theta}}/\omega_c)$. However, as such parameters cannot be null ($\tau>0$ and $\lambda>0$), implies that $\omega_c\neq \omega_\Theta$ and $\omega_c\neq \omega_{\bar{\Theta}}$, that is, there are no ``resonance states'' in the system and, therefore, the cyclotron frequency cannot coincide (``oscillate'') with the same value as the two NC frequencies (and therefore, $0\leq\omega_\Theta<\omega_c$ and $0\leq\omega_{\bar{\Theta}}<\omega_c$). Furthermore, we also note that even in the absence of the magnetic field or QHE ($B=0$) and of the Gödel-type spacetime ($\Omega=0$), the spectrum \eqref{spectra} still remains quantized (discrete) due to the presence of $\bar {\Theta}$ or $\omega_{\bar{\Theta}}$ (not of $\Theta$ or $\omega_\Theta$), i.e., such NC parameter or NC frequency acts as a type of ``NC field (or NC potential)''. Explicitly, this spectrum (100\% NC) is written as follows
\begin{equation}\label{spectra1}
E_{\pm}^{NC}=\pm\sqrt{m_0^2+2\Bar{\Theta}\bar{N}}=\pm\sqrt{m_0^2+2m_0\omega_{\Bar{\Theta}}\bar{N}},
\end{equation}
where $\bar{N}$ is a new quantum number (we make $\omega\to - \Bar{\Theta}/2m_0$), given by
\begin{equation}\label{N}
\Bar{N}\equiv\left(n+\frac{1}{2}+\frac{\Big|m_j-\frac{su\alpha}{2}\Big|-\left(m_j+\frac{su\alpha}{2}\right)}{2\alpha}\right)\geq 0.
\end{equation}

So, comparing the spectrum above with the literature, we have a generalization of the spectrum of Ref. \cite{Jellal}, that is, for $\alpha=1$ (absence of the CS) and for $\Bar{N}\to n$, we have the energy spectrum for the NC graphene.

In addition, we also note that even in the absence of the magnetic field or QHE ($B=0$) and of the NC phase space ($\Theta=\Bar{\Theta}=0$), the spectrum \eqref{spectra} still remains quantized due to the presence of $\Omega$, i.e., such vorticity parameter acts as a type of ``vorticity field''. Explicitly, this spectrum  (100\% Gödel) is written as follows (in the conical Gödel-type spacetime)
\begin{equation}\label{spectra2}
E_\pm^{G\ddot{o}del}=E_\Omega\pm\sqrt{E_\Omega^2+M_{0}^2},
\end{equation}
or better
\begin{equation}\label{spectra3}
E_\pm^{G\ddot{o}del}=2\Omega N\pm\sqrt{(2\Omega N)^2+\left(m_0-\frac{\Omega}{2}\right)^2},
\end{equation}
where $N$ is also given by \eqref{frequency}.

Besides, comparing the spectra \eqref{spectra3} with the bosonic case (spin-0 particles) without mass and in $(2+1)$-dimensions ($M=0$ and $p_z=k=0$), we verify that the fermionic case (our case) behaves in a very different way. For example, according to Ref. \cite{Carvalho}, the spectrum for such massless bosons is given by (already including the spectrum of the antiparticle)
\begin{equation}\label{spectra4}
E_\pm^{bosonic}=\Omega N_m (1\pm 1), \ \ N_m=\left(2n+1+\frac{\vert m\vert+m}{\alpha}\right),
\end{equation}
where the positive sign ($+$) is for the particle, the negative sign ($-$) is for the antiparticle, and $m=0,\pm 1,\pm 2,\ldots$ is the orbital magnetic quantum number. Already for our case, the spectrum for massless fermions is given by
\begin{equation}\label{spectra4}
E_\pm^{fermionic}=2\Omega N\pm\sqrt{(2\Omega N)^2+\left(\frac{\Omega}{2}\right)^2}.
\end{equation}

Therefore, we see that for the bosonic case, the particle has non-zero energy ($E\neq 0$) while the antiparticle has zero energy ($E=0$), which is different from the fermionic case, where both the particle and the antiparticle has non-zero energies ($E\neq 0$). In addition, in the bosonic case the spectrum depends linearly on $\Omega$, while in the fermionic case, this does not happen (due to the square root). However, for high quantum numbers $(N\gg 1)$, where $E_\pm^{fermionic}\approx 2\Omega N (1\pm 1)$, then both bosonic and fermionic cases would be a little more similar (i.e., now both spectra are linear functions on $\Omega$).

Furthermore, it is also interesting to analyze the spectrum according to the values (sign) of the quantum number $m_j$ and of the spin parameter $s$. Therefore, we obtain Table \eqref{tab1}, which shows four possible settings for the spectrum depending on the values of $m_j$ and $s$, where we define: $N_\alpha=n+\frac{1}{2}+\frac{m_j}{\alpha}$ and $N_{s}=n+\frac{1+su}{2}$, being $\epsilon^{NC}_{N_\alpha}=2m_0\omega_c^{NC}N_\alpha$ and $\epsilon^{NC}_{N_s}=2m_0\omega_c^{NC}N_s$, respectively. So, according to Table \eqref{tab1}, we see that for $m_j>0$ (settings 1 and 2) the spectra are exactly the same (have the same values) regardless of the chosen spin or spinor component ($\vert E_{\pm}^\uparrow\vert=\vert E_{\pm}^\downarrow\vert$), that is, $N$ do not depend on $s$ or $u$, however, depends on both $n$ and $m_j$ and also on $\alpha$. In this case (we have the ``maximal spectrum''), the ground state ($n=0$) still depends on $\omega^{NC}_c$ as well as $\Omega$. Already for $m_j<0$ (settings 3 and 4), we see that the spectra behave differently, i.e., the spectra are not exactly the same (do not have the same values) since now they depend on the chosen spin as well as the spinor component ($\vert E_{\pm}^\uparrow\vert\neq\vert E_{\pm}^\downarrow\vert$), that is, now $N$ depends on $s$ and $u$, however, depends only on $n$ and no longer on $m_j$ and $\alpha$. In this case (we have the ``minimal spectrum''), the ground state ($n=0$) may or not depend on $\omega^{NC}_c$ or $\Omega$ according to the value of the product $su$. Besides, in this case, it is as if the NCQHE ``lives in the flat Gödel-type spacetime'' (i.e., without the presence of a CS).
\begin{table}[h]
\centering
\begin{small}
\caption{Relativistic spectrum depend on the values of $m_j$ and $s$.} \label{tab1}
\begin{tabular}{ccc}
\hline
Setting & $(m_j,s)$ & Spectrum \\
\hline
1& $(m_j>0,+1)$ & $E_{\pm}=E_m+2\tau\Omega N_\alpha\pm \sqrt{(E_m+2\tau\Omega N_\alpha)^2-E^2_m+M_{0}^2+\epsilon^{NC}_{N_\alpha}}$\\
2& $(m_j>0,-1)$ & $E_{\pm}=E_m+2\tau\Omega N_\alpha\pm \sqrt{(E_m+2\tau\Omega N_\alpha)^2-E^2_m+M_{0}^2+\epsilon^{NC}_{N_\alpha}}$\\
3& $(m_j<0,+1)$ & $E_{\pm}=E_m+2\tau\Omega N_+\pm \sqrt{(E_m+2\tau\Omega N_+)^2-E^2_m+M_{0}^2+\epsilon^{NC}_{N_+}}$\\
4& $(m_j<0,-1)$ & $E_{\pm}=E_m+2\tau\Omega N_-\pm \sqrt{(E_m+2\tau\Omega N_-)^2-E^2_m+M_{0}^2+\epsilon^{NC}_{N_-}}$\\
\hline
\end{tabular}
\end{small}
\end{table}

Now, we need to define the restriction for the magnetic field (which will be the range allowed for $B$). As already discussed, the product $\tau\lambda$ must obey the condition: $\tau\lambda>0$. Consequently, we can obtain from this a quadratic polynomial inequality for $B$, given as follows
\begin{equation}\label{quadraticpolynomialinequality}
-B^2+B(4+\Theta\Bar{\Theta})/e\Theta-(4\Bar{\Theta}/e^2\Theta)>0,
\end{equation}
and whose solution is given by
\begin{equation}\label{solution1}
D_-<B<D_+,
\end{equation}
where
\begin{equation}\label{solution2}
D_\pm\equiv\frac{(4+\Theta\Bar{\Theta})\pm\sqrt{(4+\Theta\Bar{\Theta})^2-16\Theta\Bar{\Theta}}}{2e\Theta}.
\end{equation}

Indeed, in the absence of the NC phase space ($\Theta=\Bar{\Theta}=0$), it implies that the range of $B$ will be given by $0<B<\infty$ (or $0\leq B<\infty$), i.e., we recover the usual range of $B$ in which the QHE can manifest itself. In particular, for $\tau\lambda<0$ this does not happen, that is, we would obtain $\infty<B<0$, which is physically impossible. In that way, we see that the NC parameters restrict (limit) the values of the magnetic field, that is, for a given arbitrary value of $\Theta$ and $\Bar{\Theta}$ (and $e$), a particular range for $B$ is defined, given by the expression \eqref{solution1}.

On the other hand, it is also interesting to compare the spectra \eqref{spectra}, or the spectra from Table \eqref{tab1} (for $m_j<0$), with some other works in the literature. So, we verified that in the absence of the conical Gödel-type spacetime ($\Omega=0$ and $\alpha=1$), with $u=+1$, we obtain the spectrum of the NCQHE with AMM in Minkowski spacetime \cite{Oli6}. Already in the absence of the NC phase space ($\Theta=\Bar{\Theta}=0$), of the conical Gödel-type spacetime ($\Omega=0$ and $\alpha=1$), with $su=+1$, we obtain the spectrum of the QHE with AMM \cite{Silenko}. Furthermore, in the absence of the NC phase space ($\Theta=\Bar{\Theta}=0$), of the conical Gödel-type spacetime ($\Omega=0$ and $\alpha=1$), and of the AMM ($E_m=0$), with $N=n\geq 0$ or $N=n+1=n'\geq 1$, we obtain the spectrum of the QHE for $m_0\neq 0$ (massive case) \cite{Haldane,Schakel,Lamata,Miransky,Ishikawa2} and also for $m_0=0$ (massless case) \cite{Ishikawa2,Beneventano}. In particular, this last case with $c\to v_f\approx c/300$ (speed of light becomes the Fermi velocity) results in the ``relativistic'' Landau levels for the graphene in a commutative phase space (usual quantum phase space where $\Bar{\Theta}=0$) \cite{Novoselov,Guinea} or in a NC phase space ($\Bar{\Theta}\neq 0$) \cite{Bastos}. Therefore, we clearly see that our relativistic spectrum generalizes several particular cases in the literature.

\section{The nonrelativistic energy spectrum (nonrelativistic Landau levels)}\label{sec4}

Here, let us analyze the nonrelativistic limit (low-energy limit) of our results. To obtain this, we use a standard prescription (``cake recipe'') often used in literature to obtain the nonrelativistic limit of relativistic wave equations for massive particles (mainly for Dirac and Klein-Gordon particles) \cite{Greiner2}. So, in such prescription is necessary to consider that most of the total energy of the system stays concentrated in the rest energy of the particle \cite{Greiner2}; consequently, it implies that we can write the relativistic energy as $E\approx m_0+\varepsilon$, where $m_0 \gg \varepsilon$, $m_0 \gg E_m$, $m_0 \gg  E^{NC}_\Omega$, and $m_0 \gg \frac{(2\tau-1)\Omega}{2}$, respectively. Therefore, using this standard prescription in \eqref{spectra}, or in \eqref{dirac17}, we obtain as a result the following nonrelativistic energy spectrum (nonrelativistic Landau levels) for the NCQHE with AMM in a conical Gödel-type spacetime
\begin{equation}\label{spectra5}
\varepsilon_{n,m_j,s,u}=E^{NC}_{Zeeman-type}+\bar{E}_\Omega^{NC},
\end{equation}
where
\begin{eqnarray}\label{Zeeman}
&& E^{NC}_{Zeeman-type}=\mu_m^{NC}B=E_m+\omega_c^{NC}N,
\nonumber\\
&& \bar{E}_\Omega^{NC}=E_\Omega^{NC}-\frac{(2\tau-1)}{2}\Omega=\left(2\tau N-\frac{(2\tau-1)}{2}\right)\Omega,
\end{eqnarray}
with
\begin{eqnarray}\label{N2}
&& \mu_m^{NC}=\mu_{total}=(a+2\tau\lambda N)\mu_B, \  \ \omega_c^{NC}=\tau\lambda\omega_c,\ \omega_c=2\mu_B B,
\nonumber\\
&& N=\left(n+\frac{1}{2}+\frac{\Big|m_j-\frac{su\alpha}{2}\Big|+\left(m_j+\frac{su\alpha}{2}\right)}{2\alpha}\right)\geq 0.
\end{eqnarray}
being $E^{NC}_{Zeeman-type}$ a NC Zeeman-type spectrum, or a NC anomalous Zeeman-type spectrum (usual Zeeman spectrum modified by NC phase space and by AMM), where $\mu_m^{NC}$ is the noncommutative magnetic dipole moment (NCMDM), or the total magnetic dipole moment (TMDM) of the nonrelativistic Dirac particle (or Pauli-type particle), and $\bar{E}_\Omega^{NC}$ is a new NC Gödel spectrum (which is the ``old'' NC Gödel spectrum minus the spin-vorticity coupling term). Furthermore, it is also interesting to analyze the spectrum according to the values (sign) of the quantum number $m_j$ and of the spin parameter $s$. Therefore, we obtain Table \eqref{tab2}, which shows four possible settings for the spectrum depending on the values of $m_j$ and $s$, where $N_\alpha=n+\frac{1}{2}+\frac{m_j}{\alpha}$, and $N_{s}=n+\frac{1+su}{2}$ (we define this in the relativistic case).
\begin{table}[h]
\centering
\begin{small}
\caption{Nonrelativistic spectrum depend on the values of $m_j$ and $s$.} \label{tab2}
\begin{tabular}{ccc}
\hline
Setting & $(m_j,s)$ & Spectrum \\
\hline
1& $(m_j>0,+1)$ & $\epsilon=E_m-\frac{(2\tau-1)}{2}\Omega+(\omega_c^{NC}+2\tau\Omega)N_{\alpha}$\\
2& $(m_j>0,-1)$ & $\epsilon=E_m-\frac{(2\tau-1)}{2}\Omega+(\omega_c^{NC}+2\tau\Omega)N_{\alpha}$\\
3& $(m_j<0,+1)$ & $\epsilon=E_m-\frac{(2\tau-1)}{2}\Omega+(\omega_c^{NC}+2\tau\Omega)N_{+}$\\
4& $(m_j<0,-1)$ & $\epsilon=E_m-\frac{(2\tau-1)}{2}\Omega+(\omega_c^{NC}+2\tau\Omega)N_{-}$\\
\hline
\end{tabular}
\end{small}
\end{table}

Then, we note that the spectrum \eqref{spectra5}, or the spectra from Table \eqref{tab2}, has some similarities and differences (more similarities than differences) with the relativistic case (i.e., the relativistic particle). For example, similar to the relativistic case, the spectrum \eqref{spectra5}
\begin{itemize}
\item only admits positive energy states or positive energies ($\varepsilon>0$), whose spectrum is also for a particle with spin-1/2: spin ``up'' ($s=+1$) or spin ``down'' ($s=-1$);
\item has its degeneracy broken due to $\alpha$;
\item depends on $n$, $m_j$ and $\alpha$ for $m_j>0$ (but it does not depend on $s$ or $u$), where we
have the maximal spectrum;
\item depends on $n$, $s$ and $u$ for $m_j<0$ (but it does not depend on $m_j$ and $\alpha$), where we
have the minimal spectrum;
\item depends on the magnetic energy $E_m$, NC Gödel spectrum $E^{NC}_{\Omega}$, and of the NC cyclotron frequency $\omega_c^{NC}$;
\item does not admit resonance states ($\omega_c\neq \omega_\Theta=\frac{4}{m_0\Theta}$ and $\omega_c\neq\omega_{\bar{\Theta}}=\frac{\bar{\Theta}}{m_0}$);
\item remains quantized even in the absence of the magnetic field ($B=0$) and of the vorticity ($\Omega=0$);
\item remains quantized even in the absence of the magnetic field ($B=0$) and of the NC phase space ($\Theta=\Bar{\Theta}=0$).
\end{itemize}

However, unlike the relativistic case, the spectrum \eqref{spectra5}
\begin{itemize}
\item depends linearly on $E_m$, $E_\Omega^{NC}$, $\omega_c^{NC}$ as well as on $n$ and $m_j$ ($m_j>0$).
\end{itemize}

Now, let us compare the spectrum \eqref{spectra5},  or the spectra from Table \eqref{tab2}, with some other works in the literature. So, we verified that in the absence of the conical Gödel-type spacetime ($\Omega=0$ and $\alpha=1$), with $u=+1$, we obtain the spectrum of the NCQHE with AMM in Euclidean space \cite{Oli6}. Already in the absence of the NC phase space ($\Theta=\Bar{\Theta}=0$), of the conical Gödel-type spacetime ($\Omega=0$ and $\alpha=1$), and of the AMM ($E_m=0$), with $s=+1$ (a spinless particle) and taking $\vert m_j-1/2\vert+(m_j+1/2)\to \vert l\vert+l$ ($l=m=0,\pm 1,\pm 2,\ldots$), we obtain the usual spectrum (2D quantum harmonic oscillator-type spectrum) for the QHE \cite{Schakel,Yoshioka,Ishikawa1,Dayi,Dulat,Brand,Bhuiyan,Li}. On the other hand, in the absence of the magnetic field ($B=0$), of the NC phase space ($\Theta=\Bar{\Theta}=0$), and of the CS spacetime ($\alpha=1$), with $s=+1$, we obtain the energy spectrum in flat Gödel-type spacetime for $l<0$ \cite{Das}
and $l\geq 0$ \cite{Drukker}, i.e., there is an analogy (``equivalence'') between the flat Gödel-type metric with the Landau problem (or Landau quantization) for a particle in the plane (Cartesian or polar plane). Therefore, we clearly see that our nonrelativistic spectrum also generalizes several particular cases of the
literature.

\section{Conclusions}\label{sec5}

In this paper, we analyze the relativistic and
nonrelativistic energy spectra (fermionic Landau levels) for the noncommutative Quantum Hall effect (NCQHE) with anomalous magnetic moment (AMM) in the conical Gödel-type spacetime in $(2+1)$-dimensions, where such spacetime is the combination of the flat Gödel-type spacetime with a CS and is modeled by two parameters: $\Omega$ (vorticity parameter) and $\alpha$ (deficit angle or curvature parameter). So, to analyze these energy spectra, we start from the noncommutative Dirac equation (NCDE) with minimal and nonminimal couplings in polar coordinates, where the two coupling constants are the electric charge and the AMM of the Dirac fermion. Besides, we also consider the ``spin'' of the (2D) planar fermion, described by a parameter $s$, called the spin parameter.

Using the tetrad formalism (or spin connection), we obtain a quadratic NCDE (second-order differential equation). To solve exactly this differential equation, we apply a change of variable and then we take the asymptotic limit (behavior) of this equation. Consequently, we obtain a generalized Laguerre equation, and also a quadratic polynomial equation for the total relativistic energy. Solving this polynomial equation, we obtain as a result the relativistic energy spectrum (relativistic Landau levels) for the NCQHE with MMA in a conical Gödel-type spacetime. In particular, we verified that this spectrum (for the particle and antiparticle) explicitly depends on the quantum numbers $n$ and $m_j$ (the spectrum is quantized in terms of these quantum numbers), spin parameter $s$, spinorial parameter $u$, curvature parameter $\alpha$, magnetic energy $E_m$ (depends on the AMM and the magnetic field $B$), NC Gödel spectrum $E^{NC}_\Omega$ (depends on the vorticity parameter, NC parameter of position $\Theta$ and the magnetic field), NC cyclotron spectrum $\epsilon^{NC}_N$ (depends on the NC parameter of position and momentum $\Bar{\Theta}$ and the magnetic field, or simply on the NC cyclotron frequency $\omega_c^{NC}$), and on the NC Gödel (effective) mass $M_0$ (depends on the rest mass $m_0$, vorticity parameter, NC parameter of position and the magnetic field). Then, due to the presence of $\alpha$, the spectrum does not have a well-defined degeneracy, i.e., the function of $\alpha$ is to break the degeneracy of the spectrum. Now, due to the presence of $\Theta$ and $\Bar{\Theta}$, we were able to define two NC frequencies, given by: $\omega_{\Theta}=\frac{4}{m_0\Theta}$ and $\omega_{\Bar{\Theta}}=\frac{\Bar{\Theta}}{m_0}$, and satisfies: $0\leq \omega_{\Theta}<\omega_c$ and $0\leq\omega_{\Bar{\Theta}}<\omega_c$ ($\omega_c$ is the usual cyclotron frequency) and, therefore, there are no ``resonance states'' in the system. 

Furthermore, we also analyze the relativistic spectrum depending on the values of $m_j$ and $s$. For example, we see that for $m_j>0$, with $s=+1$ or $s=-1$, the spectrum is the same and depends on both $n$ and $m_j$ and also on $\alpha$. In other words, the spectrum does not depend on $s$ and $u$. In this case, we have the maximal spectrum, and the ground state $(n = 0)$ still depends on $\epsilon^{NC}_N$ as well as $E_\Omega^{NC}$. Already $m_j<0$, we see that the spectrum is not the same since it now depends on $s$ and $u$, and is quantized only in terms of $n$ (i.e., the spectrum does not depend on $m_j$ and $\alpha$). In this case, we have the minimal spectrum, and the ground state $(n = 0)$ may or not depend on $\epsilon^{NC}_N$ and $E_\Omega^{NC}$ according to the values of $su$. Still in this case, as the spectrum now no longer depends on $\alpha$, it is as if the NCQHE ``lives in the flat Gödel-type spacetime'', i.e., without the presence of a CS. On the other hand, we also compare the relativistic spectrum (for $m_j<0$) with some other works in the literature. For example, we verified that in the absence of the conical Gödel-type spacetime ($\Omega=0$ and $\alpha=1$), with $u=+1$, we obtain the spectrum of the NCQHE with AMM in Minkowski spacetime. Already in the absence of the NC phase space ($\Theta=\Bar{\Theta}=0$), of the conical Gödel-type spacetime, with $su=+1$, we obtain the spectrum of the QHE with AMM. Besides, in the absence of the NC phase space, of the conical Gödel-type spacetime, and of the AMM ($E_m=0$), we obtain the spectrum of the QHE for $m_0\neq 0$ (massive case) or $m_0=0$ (massless case). In particular, this last case with $c\to v_f\approx c/300$ (speed of light becomes the Fermi velocity) results in the ``relativistic'' Landau levels for the graphene in a commutative or usual quantum phase space ($\Theta=\Bar{\Theta}=0$) and in a NC phase space ($\Bar{\Theta}\neq 0$). Therefore, we clearly see that our relativistic spectrum generalizes several particular cases in the literature. 

Finally, we also analyze the nonrelativistic limit of our results. Considering that most of the total energy of the system stays concentrated in the rest energy of the particle, we obtain the nonrelativistic energy spectrum (nonrelativistic Landau levels) for the NCQHE with AMM in a conical Gödel-type spacetime. In particular, we verified that this spectrum is written in terms of two spectra: the NC Zeeman-type spectrum (depends on $E_m$ and $\omega^{NC}_c$, or of the NC total magnetic dipole moment $\mu_m^{NC}$ ), and new NC Gödel spectrum $\Bar{E}^{NC}_\Omega$ (depends on the ``old'' NC Gödel spectrum $E^{NC}_\Omega$ and of the spin-vorticity coupling term). Then, we note that the nonrelativistic spectrum has some similarities and differences with the relativistic case. For example, similar to the relativistic case, the nonrelativistic spectrum only admits positive energies, whose spectrum is also for a particle with spin-1/2; has its degeneracy broken due to $\alpha$; depends on $n$, $m_j$ and $\alpha$ for $m_j>0$; depends on $n$, $s$ and $u$ for $m_j<0$; does not admit resonance states; and remains quantized even in the absence of the magnetic field and of the vorticity or of the NC phase space. However, unlike the relativistic case, the nonrelativistic spectrum depends linearly on $E_m$, $E^{NC}_\Omega$, $\omega^{NC}_c$ as well as on $n$ and $m_j$. In addition, we also compare the nonrelativistic spectrum with some other works in the literature, where we verified that in the absence of the conical Gödel-type spacetime, we obtain the spectrum of the NCQHE with AMM in Euclidean space. Already in the absence of the NC phase space, of the conical Gödel-type spacetime, and of the AMM, we obtain the usual spectrum for the QHE. On the other hand, in the absence of the magnetic field, of the NC phase space, and of the CS spacetime, we obtain the energy spectrum in flat Gödel-type spacetime. Therefore, we clearly see that our nonrelativistic spectrum also generalizes several particular cases in the literature.

\section*{Acknowledgments}

\hspace{0.5cm}The author would like to thank the Conselho Nacional de Desenvolvimento Cient\'{\i}fico e Tecnol\'{o}gico (CNPq) for financial support.

\section*{Data availability statement}

\hspace{0.5cm} This manuscript has no associated data or the data will not be deposited. [Author’s comment: There is no data because this is theoretical work based on calculations to obtain the relativistic and nonrelativistic Landau levels for the noncommutative quantum Hall effect with anomalous magnetic moment in a conical Gödel-type spacetime.]

\end{document}